\documentclass{article}
\usepackage{graphicx}
\usepackage{epsfig}

\def\spose#1{\hbox to 0pt{#1\hss}}
\def\simlt{\mathrel{\spose{\lower 3pt\hbox{$\mathchar"218$}}
     \raise 2.0pt\hbox{$\mathchar"13C$}}}
\def\simgt{\mathrel{\spose{\lower 3pt\hbox{$\mathchar"218$}}
     \raise 2.0pt\hbox{$\mathchar"13E$}}}
\begin{document}

\noindent
{\LARGE INTRODUCTION}

Martin J Rees\\
{\it Institute of Astronomy, Madingley Road, Cambridge, CB3 0HA}


\begin{abstract}
 It is embarrassing that 95\% of the universe is unaccounted for. 
Galaxies and  larger-scale cosmic structures  are composed mainly of  `dark 
matter'  whose nature is still unknown. Favoured candidates  are 
weakly-interacting particles that have survived from the very early 
universe, but more exotic options  cannot be excluded. (There are strong 
arguments that the dark matter  is not composed of baryons).  Intensive 
experimental searches  are being made for the `dark'  particles (which 
pervade our entire galaxy), but we have indirect clues to their nature too. 
Inferences   from galactic dynamics and gravitational lensing allow 
astronomers to `map' the dark matter distribution; comparison with 
numerical simulations of galaxy formation can constrain (eg) the particle 
velocities and collision cross sections. And, of course, progress in 
understanding the extreme  physics of the ultra-early universe could offer 
clues to what particle might  have existed then, and how many would have 
survived.

   The mean cosmic density of dark matter (plus baryons) is  now pinned 
down  to be only about 30\% of the so-called  critical density 
corresponding to a `flat' universe. However, other recent  evidence -- 
microwave background anisotropies, complemented by data on distant 
supernovae -- reveals  that our universe actually is `flat', but that its 
dominant ingredient  (about 70\% of the total mass-energy) is 
something quite unexpected ---  `dark energy' pervading all space,  with 
negative pressure.   We now confront two mysteries:

  (i) Why does the universe have three quite distinct  basic ingredients -- 
baryons, dark matter and dark energy -- in the proportions (roughly)  5\%, 
25\% and 70\%?

  (ii) What are the (almost certainly profound)  implications of the  `dark 
energy' for fundamental physics?
\end{abstract}

\section {SOME HISTORY}

      Astronomers have  long known that  galaxies and clusters  would fly 
apart unless they were  held together by the gravitational pull of much 
more material than we actually see.

   The strength of the case built up gradually. The argument that clusters 
of galaxies would be unbound without dark matter dates back to  
Zwicky (1937)  and others in the 1930s.  Kahn and Woltjer (1959) pointed 
out that the motion of Andromeda towards us implied  that there must be 
dark matter in our  Local Group of galaxies. But the  dynamical evidence 
for massive halos (or `coronae') around individual  galaxies firmed up 
rather later (e.g.  Roberts and Rots 1973, Rubin, Thonnard and Ford 1978).

   Two 1974 papers were  specially influential in the latter context. Here 
is a quote from each:

\begin{quote}
   The mass of galactic coronas exceeds the mass of populations of known 
stars by one order of magnitude, as do the effective dimensions. .... The 
mass/luminosity ratio rises to f=100 for spiral and $f=120$ for elliptical 
galaxies. With $H = 50$ km/sec/Mpc this ratio for the Coma cluster is 170
(Einasto, Kaasik and Saar 1974)
\end{quote}

\begin{quote}
 Currently-available observations strongly indicate that the mass of 
spiral galaxies increases almost linearly with radius to nearly 1 Mpc.... 
and that the ratio of this mass to the  light within the Holmberg radios, 
$f$, is  200 ($M/L_\odot$). (Ostriker, Peebles and Yahil, 1974).
\end{quote}

  The  amount of dark matter, and how it is distributed, is now far better 
established  than it was  when those papers were written.  The immense advances in 
delineated dark matter in clusters and in individual galaxies  are manifest 
in the programme for this meeting. The rapid current progress stems from 
the confluence of several  new kinds of data within the same few-year 
interval: optical surveys  of large areas and high redshifts, CMB 
fluctuation measurements, sharp X-ray images, and so forth.

     The progress has not been solely observational. Over the last 20 
years, a compelling  theoretical perspective for the emergence of cosmic 
structure has been developed.  The expanding universe  is unstable to the 
growth of structure, in the sense that regions that start off 
{\it very slightly} 
overdense have their expansion slowed by their excess gravity, and evolve 
into conspicuous density contrasts.  According to this `cold dark matter' 
(CDM) model,  the present-day  structure of galaxies and clusters is 
moulded by the gravitational aggregation of non-baryonic matter, which is 
an essential ingredient of the early universe (Pagels and Primack 1982, 
Peebles 1982,  Blumenthal {\it et al.} 1984, Davis {\it et al.} 1985).     These models 
have been firmed up by  vastly improved simulations,  rendered possible by 
burgeoning  computer power. And astronomers can now compare these `virtual 
universes' with the real one, not just at the present era but (by observing 
very distant objects)  can probe back towards the formative stages when the 
first galaxies  emerged.

     The following comments are intended to provide a context for the later 
papers. (For that reason, I do not give detailed references to the topics 
covered by other speakers -- just some citations of historical interest).

\section{THE CASE FOR DARK MATTER}

\subsection{Baryons}

The inventory of cosmic baryons is readily compiled. Stars and their 
remnants, and gas in galaxies, contribute no more than 1\% of the 
critical density (i.e. they give $\Omega_b   < 0.01$). However  several percent  more 
could be contributed by diffuse material pervading  intergalactic space: 
warm gas (with $kT \simeq 0.1$ keV) in groups of galaxies and loose clusters, and cooler gas 
pervading intergalactic space that  manifests itself via the `picket fence' 
absorption lines in quasar spectra. (Rich clusters are rare, so their 
conspicuous gas content, at several KeV, is not directly significant for 
the total inventory,  despite its  importance as a probe)

  These baryon estimates are concordant with those inferred by matching 
the He and D abundances at the birth of galaxies with the predicted outcome 
of  nucleosynthesis in the big bang, which is sensitive to the primordial 
baryon/photon ratio, and thus to   $\Omega_b$.  The  observational estimates have 
firmed up, with improved measurements of  deuterium  in high-$z$ 
absorbing clouds. The best fit occurs for $\Omega_b \simeq 0.02h^{-2}$ where $h$ is the Hubble constant in units of 100 km s$^{-1}$ Mpc$^{-1}$. Observations favour $h \simeq 0.7$. 

       $\Omega_b$  is now pinned down by  a variety of argument to be 
$0.04 - 0.05$. This corresponds to only $\sim 0.3$ baryons per cubic metre, a value so low that it leaves little scope for dark  baryons. (It is 
therefore unsurprising that the MACHO/OGLE  searches should have found that 
compact objects do not make a substantial  contribution to the total  mass 
of our own galactic halo.)

\subsection{How much dark matter?}
  An important recent development is that $\Omega_{DM}$ can now be constrained to a 
value  around 0.25 by several independent lines of evidence:

  (i)   One of the most ingenious and convincing arguments  comes from 
noting that baryonic matter in clusters -- in galaxies, and in intracluster 
gas -- amounts to  $0.15-0.2$ of the inferred virial mass (White {\it et al.} 1993) 
. If clusters were a fair sample of the universe, this would then be 
essentially the same as the cosmic ratio of baryonic to total mass.    Such 
an argument could not be applied to an individual galaxy,  because  baryons 
segregate towards the centre. However, there is no such segregation on the 
much larger scale of clusters: only a small correction is necessary to 
allow for  baryons expelled during the cluster formation process.

  (ii)   Very distant galaxies appear  distorted, owing to gravitational 
lensing by intervening galaxies and clusters.   Detailed modelling of  the 
mass-distributions needed to cause the observed distortions yields a 
similar estimate. This is a straight measurement of $\Omega_{DM}$ which (unlike (i)) 
does not involve assumptions about $\Omega_b$, though it does depend on having 
an accurate measure of the clustering amplitude.

    (iii) Another argument is based on the way density contrasts grow
during the cosmic expansion: in a low density universe, the expansion
kinetic energy overwhelms gravity, and the growth of structure
saturates at recent epochs. The existence of conspicuous clusters of
galaxies with redshifts as large as $z=1$ is hard to reconcile with the
rapid recent growth of structure that would be expected if $\Omega_{DM}$ were
unity. More generally, numerical simulations based on the cold dark
matter (CDM) model model are a better fit to the present-day structure
for this value of $\Omega_{DM}$(partly because the initial fluctuation
spectrum has too little long-wavelength power if $\Omega_{DM}$ is unity).

    Other methods will soon offer independent estimates. For instance,  
$\Omega_{DM}$
can be estimated from  the  deviations from the Hubble flow induced by 
large-scale irregularities in the  mass distribution on supercluster scales.

\subsection{What could the dark matter be?}

   The  dark matter  is not  primarily baryonic. The amount of deuterium 
calculated to emerge from the big bang  would  be  far lower than observed 
if the average  baryon density  were $\sim 2$ (rather than $\sim 0.3$)   per cubic metre. 
Extra exotic particles that do not participate in nuclear reactions, 
however, would not scupper the concordance.

    Beyond the negative statement that it is non-baryonic, the nature of 
the dark matter  still eludes us.  This key question may yield to a 
three-pronged attack:

1. $\underline{\hbox {Direct detection}}$. As described by other contributors to this meeting, 
several groups are developing cryogenic detectors for supersymmetric 
particles and axions This is an exciting quest. Of course, not even 
optimists can be confident that the actual dark matter particles have 
parameters within the range that these experiments are yet sensitive to. 
But the stakes are high: detection of most of the gravitating stuff in 
the universe, as well as a new class of elementary particle. So it seems 
well worth committing to these experiments  funding that is equivalent to 
a small fraction of the cost of a major accelerator

 2. $\underline{\hbox {Progress in particle physics}}$. Important recent
measurements suggest that neutrinos have non-zero masses; this result
has crucially important implications for physics beyond the standard
model. The inferred neutrino masses seem, however, too low to be cosmologically
important.  If the masses and cross-sections of supersymmetric
particles were known, it should be possible to predict how many
survive, and their contribution to $\Omega$, with the same confidence
with which we can compute the nuclear reactions that control
primordial nucleosynthesis. Associated with such progress, we might
expect a better understanding of how the baryon-antibaryon asymmetry
arose, and the consequence for $\Omega_b$.  Optimists may hope for
progress on still more exotic options.

3. $\underline{\hbox {Simulations of galaxy formation and large-scale
structure}}$.  When and how galaxies form, the way they are clustered,
and the density profiles within individual systems, depend on what
their gravitationally-dominant constituent is. A combination of better
data and better simulations is starting to set generic constraints on
the options. The CDM model works well. But there are claimed
discrepancies, though many of us suspect these may ease when the galaxy formation process is better understood. For instance the centre of a halo would, according to the
simulations, have a `cusp' rather than the measured uniform-density core: this
discrepancy has led some authors to explore modifications where the
particles are assumed to have significant collision probabilities, or to
be moving with non-negligible velocities (i.e. `warm' not
cold.). These calculations are in any case offering interesting constraints on  the properties of heavy
supersymmetric particles. (Also, straight astronomical observations
can rule out a contribution to $\Omega$  of more than 0.01 from neutrinos
-- this is compatible with current experimental estimates.)

\section{DARK ENERGY}

  The inference that our universe is dominated by dark matter is in itself 
a discovery of the first magnitude. But the realisation that  even more 
mass-energy is in some still  more mysterious form -- dark energy latent in 
space itself -- came as a surprise, and probably has even greater import 
for fundamental physics.

 If this meeting had been taking place 3 years ago, the more open-minded 
among us would have given equal billing to two options: a hyperbolic 
universe with   $\Omega$  of 0.3, (in which it would  be a coincidence that the 
Robertson-Walker curvature radius was comparable with the present Hubble radius), 
or a flat universe in which something other than CDM makes up the balance, 
equivalent to  $\Omega$  of 0.7 (In this case it would be a coincidence that two 
quite different invisible substances make comparable contributions).

       But  it is now clear that only the second option remains in the 
running: there is compelling  evidence that the universe is flat.  This 
evidence comes from the  slight temperature-differences over the sky in the 
background radiation, due to density irregularities which are the 
precursors of cosmic structure. Theory tell us that the temperature 
fluctuations  should  be biggest on a particular length scale that is 
related to the distance a sound wave can travel in the early universe. 
The angular scale corresponding to this length depends, however, on the 
geometry of the universe. If dark matter and baryons  were all, we wouldn't 
be in a flat universe -- the geometry would be hyperbolic.  Distant objects 
would look smaller than in a flat  universe.  In 2001-02, measurements from 
balloons and from Antarctica  pinned down the angular scale of this 
`doppler peak': the results   indicated  `flatness' -- a result now 
confirmed with greater precision by the WMAP satellite.

    A value of  0.3 for $\Omega_{DM}$ would imply (were  there no other energy in the 
universe) an angle smaller by almost a factor of 2 -- definitely in 
conflict with observations.   So what's the  other 70\%? It is not 
dark matter but  something   that does not cluster -- some energy latent in 
space.  The simplest form of this  idea goes back to 1917 when Einstein 
introduced the cosmological constant, or  lambda. A positive lambda can be 
interpreted, in the context of the ordinary Friedman equations, as a fixed 
positive energy density in all space.  This leads to a repulsion because, 
according to Einstein's equation, gravity depends on pressure as well as 
density, and vacuum energy has such a large negative pressure -- tension 
-- that the  net effect is repulsive.

        Einstein's cosmological constant is just one of the options. A 
class of more general models is being explored (under names such as 
`quintessence')   where the energy is time-dependent. Any form of dark 
energy must have negative pressure  to be compatible with observations -- 
unclustered relativistic particles, for instance, can be ruled out as 
candidates. The argument is straightforward: at present,  dark energy 
dominates the universe --  it amounts to around 70\% of the total 
mass-energy. But had it  been equally dominant in the past,  it would 
have inhibited the growth of the density contrasts in cosmic structures, 
which occurred gravitational instability.  This is because the growth timescale for gravitational instability is $\sim\left(G\rho_c\right)^{-{1 \over 2}}$, where $\rho_c$ is the density of the component that participates in the clustering, whereas the expansion timescale scales as $\left(G\rho_{total}\right)^{-{1 \over 2}}$ when curvature is unimportant. If $\rho_{total}$ exceeds $\rho_c$, the expansion is faster, so the growth is impeded.  (Meszaros, 1974) 

     In the standard model, density contrasts in the dark matter grow by 
nearly 1000 since recombination. If this growth had been suppressed, the 
existence of  present-day clusters would therefore require irregularities 
that were already of substantial amplitude  at the recombination epoch, 
contrary to the evidence from CMB fluctuations. For the `dark energy' to be 
less dominant in the past, its density must depend on the scale factor $R$ more slowly than 
the  $R^{-3}$  dependence of pressure-free matter -- i.e. its PdV work must be 
negative.  Cosmologists have introduced a parameter $w$ such that 
$p = w\rho   c^2$. 
A more detailed treatment    yields the  requirement that 
$w < -0.5$. 
This comes from taking account of baryons and dark matter, and requiring 
 that   dark energy  should not have inhibited  the growth of structure 
so  much  that it destroyed the concordance between the CMB fluctuations 
(which measure the amplitude at recombination) and the present-day 
inhomogeneity. Note however that unless its value  is -1 (the special case 
of a classical cosmological constant)  $w$ will generally be time-dependent. 
In principle $w(t)$ can be pinned down  by measuring the Hubble expansion 
rate at different redshifts

   This line of argument would in itself have led to a prediction of 
accelerating cosmic expansion.  However, as it turned out, studies of the 
redshift versus the apparent brightness of  distant SNIa -- strongly 
suggestive if not yet completely compelling --  had  already  conditioned 
us to  the belief that galaxies are indeed  dispersing at an accelerating 
rate. As often in science, a clear picture gradually builds up, but the 
order in which the bits of the jigsaw fall into place  is a matter of 
accident or contingency.  
   CMB fluctuations  alone can now  pin down $\Omega_{DM}$   and the curvature 
independent of all the other measurements. 

      The `modern' interest in the cosmological constant  stems from its 
interpretation as a vacuum energy. This leads to the reverse problem: Why 
is lambda at least 120 powers of 10 smaller than its `natural' value, even 
though the effective vacuum density must have been very high in order to 
drive inflation. If lambda is fully resurrected,  it will be a posthumous 
`coup' for de Sitter. His model, dating from the 1920s, not only describes 
inflation, but would then also describe future aeons of our cosmos with 
increasing accuracy. Only for the 50-odd decades of logarithmic time 
between the end of inflation and the present would it need modification!. 
But of course the dark energy could have a more complicated  and 
time-dependent nature -- though  it must have negative pressure, and  it 
must not participate in gravitational clustering.

\section{SUMMARY AND PROSPECTS}

      Cosmologists can now   proclaim with confidence (but with some 
surprise too)  that, in round numbers, our universe consists of 5\% 
baryons, 25\%  dark matter, and 70\% dark energy.  It is indeed 
embarrassing that 95\% of the universe is unaccounted for: even the 
dark matter is of quite uncertain nature, and the dark energy is a complete 
mystery.

     The network of key arguments is summarised in Figure 1.  Historically, 
the  supernova evidence came first. But had the order of events been 
different, one could have predicted an acceleration on the basis of CDM 
evidence alone; the supernovae would then have offered gratifying 
corroboration (despite the unease about possible poorly-understood 
evolutionary effects).

 Our universe is flat, but with a strange mix of ingredients. Why should 
these all give comparable contributions (within a modest factor) when they 
could have differed by a hundred powers of ten?

    In the coming decade, we can expect advances on several fronts.
       Physicists may well develop  clearer ideas on what determined the 
favouritism for matter over antimatter in the early universe, and on the 
particles that make up the dark matter.   Understanding the dark 
energy, and indeed the big bang itself, is perhaps a remoter goal, but ten years from now  
theorists may well  have  replaced the boisterous variety of ideas on the 
ultra-early universe by a firmer best buy. They will do this by discovering 
internal inconsistencies in some contending  theories, and thereby 
narrowing down the field. Better still, maybe one theory will earn 
credibility by explaining things we can observe, so that we can apply it 
confidently even to things we cannot directly observe.  In consequence, we 
may have a better insight into the origin of the fluctuations, the   dark 
energy,  and perhaps the big bang itself.

Inflation models have two  generic expectations; that the universe should 
be flat and that the fluctuations should be gaussian and adiabatic (the 
latter because baryogenesis would occur at a later stage than inflation).
   But other features of the  fluctuations  are in principle measurable and 
would be a diagnostic of the specific  physics. One, the ratio of the 
tensor and scalar amplitudes of the fluctuations,  will have to await the 
next generation of CMB experiments, able to probe the polarization on small 
angular scales.  Another discriminant among different theories is  the 
extent to which the fluctuations deviate from a Harrison-Zeldovich 
scale-independent format ($n=1$ in the usual notation); they could follow a 
different power law (i.e. be tilted) , or have a `rollover' so that the 
spectral slope is itself a function of scale. Such effects are  already 
being constrained by WMAP data, in combination with evidence on smaller 
scales from present-day clustering, from the statistics of the Lyman alpha 
absorption-line `forest' in quasar spectra, and from indirect evidence on 
when the first minihalos collapsed, signalling the formation of the first 
Population III stars that ended the cosmic dark age.

      In parallel, there will be progress in `environmental
cosmology'.  The new generation of 10-metre class ground based
telescopes will give more data on the universe at earlier cosmic
epochs, as well as better information on gravitational lensing by dark
matter. And there will be progress by theorists too.  The behaviour of
the dark matter, if influenced solely by gravity, can already be
simulated with sufficient accuracy. Gas dynamics, including shocks and
radiative cooling, can be included too (though of course the
resolution isn't adequate to model turbulence, nor the viscosity in
shear layers).  Spectacular recent simulations have been able to
follow the formation of the first stars. But the later stages of
galactic evolution, where feedback is important, cannot be modelled
without parametrising such processes in a fashion guided by physical
intuition and observations. Fortunately, we can expect rapid
improvements, from observations in all wavebands, in our knowledge of
galaxies, and the high-redshift universe.

 Via a combination of improved observations, and ever more refined 
simulations, we can hope to elucidate how our elaborately structured cosmos 
emerged from a near-homogeneous   early universe.

\noindent
References

\noindent
Blumenthal, G, Faber, S,  Primack, J.R, and Rees, M.J. 1984 Nature 311, 517\par
\noindent
Davis, M, Efstathiou, G.P, Frenk, C.S. and White, S.D.M., 1985 Astrophys. J. 292, 371\par
\noindent
Einasto, J , Kaasik, A  and Saar, E, 1974  Nature 250, 309\par
\noindent
Kahn, F and Woltjer, L, 1959 Astrophys. J. 130, 705\par
\noindent
Meszaros, P. 1974 Astr. Astrophys. 37, 225\par
\noindent
Ostriker, J. Peebles, P.J.E., and Yahil, A, 1974  Astrophys.J (Lett) 193, L1\par
\noindent
Pagels, H. and Primack, J.R. 1982 Phys. Rev. Lett. 48, 223\par
\noindent
Peebles, P.J.E. 1982 Astrophys.J. (Lett) 263, L1\par
\noindent
Roberts,  M.S. and Rots, A.H. 1973 Astr.Astrophys 26, 483.\par
\noindent
Rubin, V.C., Thonnard, N., and Ford, W.K., 1978 Astrophys J. (Lett) 225 , 
L107\par
\noindent
White, S.D.M., Navarro, J.F., Evrard, A.E. and Frenk, C.S. 1993, Nature 366, 429\par
\noindent
Zwicky, F, 1937  Astrophys. J  86, 217\par
\vfill\eject
\begin{figure}
\epsfig{file=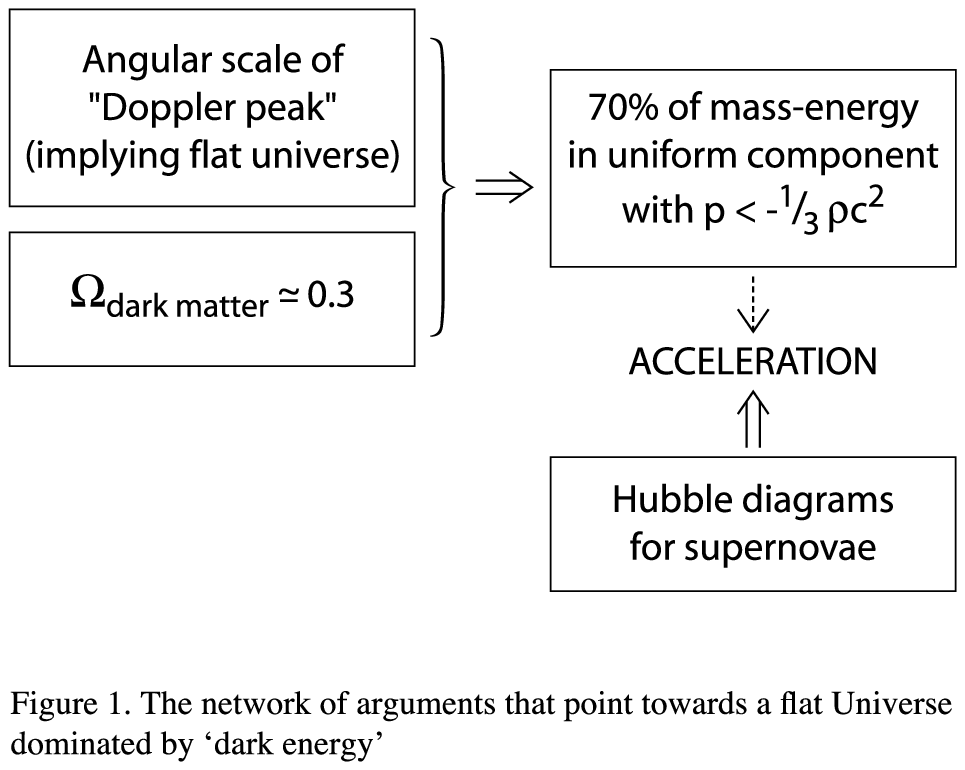, width=16cm}
\end{figure}
\end{document}